
\newcount\mgnf\newcount\tipi\newcount\tipoformule\newcount\driver
\newcount\indice
\driver=1        
\mgnf=0          
\tipi=2          
\tipoformule=0   
\indice=1        

\ifnum\mgnf=0
   \magnification=\magstep0
   \hsize=14truecm\vsize=20.truecm
   \parindent=0.3cm\baselineskip=0.45cm\fi
\ifnum\mgnf=1
   \magnification=\magstep1\hoffset=0.truecm
   \hsize=14truecm\vsize=20truecm
   \baselineskip=18truept plus0.1pt minus0.1pt \parindent=0.9truecm
   \lineskip=0.5truecm\lineskiplimit=0.1pt      \parskip=0.1pt plus1pt\fi
\let\a=\alpha \let\b=\beta       \let\d=\delta  \let\e=\varepsilon
\let\z=\zeta  \let\h=\eta   \let\th=\vartheta    \let\l=\lambda
\let\m=\mu        \let\x=\xi              
\let\s=\sigma \let\t=\tau   \let\f=\varphi

 \let\D=\Delta     \let\L=\Lambda  
     \let\F=\Phi


\global\newcount\numsec\global\newcount\numfor
\global\newcount\numapp\global\newcount\numcap
\global\newcount\numfig\global\newcount\numpag
\global\newcount\numnf

\def\SIA #1,#2,#3 {\senondefinito{#1#2}%
\expandafter\xdef\csname #1#2\endcsname{#3}\else
\write16{???? ma #1,#2 e' gia' stato definito !!!!} \fi}

\def \FU(#1)#2{\SIA fu,#1,#2 }

\def\etichetta(#1){(\veroparagrafo.\veraformula)%
\SIA e,#1,(\veroparagrafo.\veraformula) %
\global\advance\numfor by 1%
\write15{\string\FU (#1){\equ(#1)}}%
\write16{ EQ #1 ==> \equ(#1)  }}
\def\etichettaa(#1){(A\veraappendice.\veraformula)
 \SIA e,#1,(A\veraappendice.\veraformula)
 \global\advance\numfor by 1
 \write15{\string\FU (#1){\equ(#1)}}
 \write16{ EQ #1 ==> \equ(#1) }}
\def\getichetta(#1){Fig. \verafigura
 \SIA g,#1,{\verafigura}
 \global\advance\numfig by 1
 \write15{\string\FU (#1){\graf(#1)}}
 \write16{ Fig. #1 ==> \graf(#1) }}
\def\retichetta(#1){\numpag=\pgn\SIA r,#1,{\verapagina}
 \write15{\string\FU (#1){\rif(#1)}}
 \write16{\rif(#1) ha simbolo  #1  }}
\def\etichettan(#1){(n\verocapitolo.\veranformula)
 \SIA e,#1,(n\verocapitolo.\veranformula)
 \global\advance\numnf by 1
\write16{\equ(#1) <= #1  }}

\newdimen\gwidth
\gdef\profonditastruttura{\dp\strutbox}
\def\senondefinito#1{\expandafter\ifx\csname#1\endcsname\relax}
\def\BOZZA{
\def\alato(##1){
 {\vtop to \profonditastruttura{\baselineskip
 \profonditastruttura\vss
 \rlap{\kern-\hsize\kern-1.2truecm{$\scriptstyle##1$}}}}}
\def\galato(##1){ \gwidth=\hsize \divide\gwidth by 2
 {\vtop to \profonditastruttura{\baselineskip
 \profonditastruttura\vss
 \rlap{\kern-\gwidth\kern-1.2truecm{$\scriptstyle##1$}}}}}
\def\verapagina{
{\romannumeral\number\numcap}.\number\numsec.\number\numpag}}

\def\alato(#1){}
\def\galato(#1){}
\def\veroparagrafo{\number\numsec}\def\veraformula{\number\numfor}
\def\veraappendice{\number\numapp}
\def\verapagina{\number\pageno}\def\veranformula{\number\numnf}
\def\verafigura{{\romannumeral\number\numcap}.\number\numfig}
\def\verocapitolo{\number\numcap}\def\veranformula{\number\numnf}
\def\Eqn(#1){\eqno{\etichettan(#1)\alato(#1)}}
\def\eqn(#1){\etichettan(#1)\alato(#1)}

\def\Eq(#1){\eqno{\etichetta(#1)\alato(#1)}}
\def\eq(#1){\etichetta(#1)\alato(#1)}
\def\Eqa(#1){\eqno{\etichettaa(#1)\alato(#1)}}
\def\eqa(#1){\etichettaa(#1)\alato(#1)}
\def\dgraf(#1){\getichetta(#1)\galato(#1)}
\def\drif(#1){\retichetta(#1)}

\def\eqv(#1){\senondefinito{fu#1}$\clubsuit$#1\else\csname fu#1\endcsname\fi}
\def\equ(#1){\senondefinito{e#1}\eqv(#1)\else\csname e#1\endcsname\fi}
\def\graf(#1){\senondefinito{g#1}\eqv(#1)\else\csname g#1\endcsname\fi}
\def\rif(#1){\senondefinito{r#1}\eqv(#1)\else\csname r#1\endcsname\fi}

\openin14=\jobname.aux \ifeof14 \relax \else
\input \jobname.aux \closein14 \fi



\ifnum\tipoformule=1\let\Eq=\eqno\def\eq{}\let\Eqa=\eqno\def\eqa{}
\def\equ{}\fi


{\count255=\time\divide\count255 by 60 \xdef\hourmin{\number\count255}
	\multiply\count255 by-60\advance\count255 by\time
   \xdef\hourmin{\hourmin:\ifnum\count255<10 0\fi\the\count255}}

\def\oramin{\hourmin }

\def\data{\number\day/\ifcase\month\or gennaio \or febbraio \or marzo \or
aprile \or maggio \or giugno \or luglio \or agosto \or settembre
\or ottobre \or novembre \or dicembre \fi/\number\year;\ \oramin}

\setbox200\hbox{$\scriptscriptstyle \data $}

\newcount\pgn \pgn=1
\def\foglio{\number\numsec:\number\pgn
\global\advance\pgn by 1}
\def\foglioa{A\number\numsec:\number\pgn
\global\advance\pgn by 1}

%

\footline={\rlap{\hbox{\copy200}}}

%
%
\newdimen\xshift \newdimen\xwidth
%
%
\def\ins#1#2#3{\vbox to0pt{\kern-#2 \hbox{\kern#1 #3}\vss}\nointerlineskip}
%
%
\def\insertplot#1#2#3#4{
    \par \xwidth=#1 \xshift=\hsize \advance\xshift
     by-\xwidth \divide\xshift by 2 \vbox{
  \line{} \hbox{ \hskip\xshift  \vbox to #2{\vfil
 \ifnum\driver=0 #3  
                 \special{ps: plotfile #4.ps} 
 \fi
 \ifnum\driver=1  #3    \includegraphics{#4.ps}       \fi
 \ifnum\driver=2  #3   \ifnum\mgnf=0
                       \special{#4.ps 1. 1. scale}\fi
                       \ifnum\mgnf=1
                       \special{#4.ps 1.2 1.2 scale}\fi\fi
 \ifnum\driver=5  #3   \fi}
\hfil}}}

\newdimen\xshift \newdimen\xwidth \newdimen\yshift
\def\eqfig#1#2#3#4#5{
\par\xwidth=#1 \xshift=\hsize \advance\xshift
by-\xwidth \divide\xshift by 2
\yshift=#2 \divide\yshift by 2
\line{\hglue\xshift \vbox to #2{\vfil
\ifnum\driver=0 #3
\special{ps: plotfile #4.ps} 
\fi
\ifnum\driver=1 #3 \includegraphics{#4.ps}\fi
\ifnum\driver=2 #3 \special{
\ifnum\mgnf=0 #4.ps 1. 1. scale \fi
\ifnum\mgnf=1 #4.ps 1.2 1.2 scale\fi}
\fi}\hfill\raise\yshift\hbox{#5}}}

\def\figini#1{
\def\8{\write13}
\message{************************************************************}
\message{Sto disegnando la fig. #1}
\message{************************************************************}
\catcode`\%=12\catcode`\{=12\catcode`\}=12
\catcode`\<=1\catcode`\>=2
\openout13=#1.ps}

\def\figfin{
\closeout13
\catcode`\%=14\catcode`\{=1
\catcode`\}=2\catcode`\<=12\catcode`\>=12
\message{l' ho disegnata!}
\message{************************************************************}
}

\newskip\ttglue
\def\TIPIO{
\font\setterm=amr7 
\font\settesy=amsy7 \font\settebf=ambx7 
\def \settepunti{\def\rm{\fam0\setterm}
\textfont0=\setterm   
\textfont2=\settesy   
\textfont\bffam=\settebf  \def\bf{\fam\bffam\settebf}
\normalbaselineskip=9pt\normalbaselines\rm
}\let\nota=\settepunti}
\def\annota#1#2{\footnote{${}^#1$}{\nota#2\vfill}}

\def\TIPITOT{
\font\twelverm=cmr12
\font\twelvei=cmmi12
\font\twelvesy=cmsy10 scaled\magstep1
\font\twelveex=cmex10 scaled\magstep1
\font\dodici=cmbx10 scaled\magstep1
\font\twelveit=cmti12
\font\twelvett=cmtt12
\font\twelvebf=cmbx12
\font\twelvesl=cmsl12
\font\ninerm=cmr9
\font\ninesy=cmsy9
\font\eightrm=cmr8
\font\eighti=cmmi8
\font\eightsy=cmsy8
\font\eightbf=cmbx8
\font\eighttt=cmtt8
\font\eightsl=cmsl8
\font\eightit=cmti8
\font\sixrm=cmr6
\font\sixbf=cmbx6
\font\sixi=cmmi6
\font\sixsy=cmsy6
\font\twelvetruecmr=cmr10 scaled\magstep1
\font\twelvetruecmsy=cmsy10 scaled\magstep1
\font\tentruecmr=cmr10
\font\tentruecmsy=cmsy10
\font\eighttruecmr=cmr8
\font\eighttruecmsy=cmsy8
\font\seventruecmr=cmr7
\font\seventruecmsy=cmsy7
\font\sixtruecmr=cmr6
\font\sixtruecmsy=cmsy6
\font\fivetruecmr=cmr5
\font\fivetruecmsy=cmsy5
\textfont\truecmr=\tentruecmr
\scriptfont\truecmr=\seventruecmr
\scriptscriptfont\truecmr=\fivetruecmr
\textfont\truecmsy=\tentruecmsy
\scriptfont\truecmsy=\seventruecmsy
\scriptscriptfont\truecmr=\fivetruecmr
\scriptscriptfont\truecmsy=\fivetruecmsy
\def \eightpoint{\def\rm{\fam0\eightrm}
\textfont0=\eightrm \scriptfont0=\sixrm \scriptscriptfont0=\fiverm
\textfont1=\eighti \scriptfont1=\sixi   \scriptscriptfont1=\fivei
\textfont2=\eightsy \scriptfont2=\sixsy   \scriptscriptfont2=\fivesy
\textfont3=\tenex \scriptfont3=\tenex   \scriptscriptfont3=\tenex
\textfont\itfam=\eightit  \def\it{\fam\itfam\eightit}%
\textfont\slfam=\eightsl  \def\sl{\fam\slfam\eightsl}%
\textfont\ttfam=\eighttt  \def\tt{\fam\ttfam\eighttt}%
\textfont\bffam=\eightbf  \scriptfont\bffam=\sixbf
\scriptscriptfont\bffam=\fivebf  \def\bf{\fam\bffam\eightbf}%
\tt \ttglue=.5em plus.25em minus.15em
\setbox\strutbox=\hbox{\vrule height7pt depth2pt width0pt}%
\normalbaselineskip=9pt
\let\sc=\sixrm  \normalbaselines\rm
\textfont\truecmr=\eighttruecmr
\scriptfont\truecmr=\sixtruecmr
\scriptscriptfont\truecmr=\fivetruecmr
\textfont\truecmsy=\eighttruecmsy
\scriptfont\truecmsy=\sixtruecmsy
}\let\nota=\eightpoint}

\newfam\msbfam   
\newfam\truecmr  
\newfam\truecmsy 
\newskip\ttglue
\ifnum\tipi=0\TIPIO \else\ifnum\tipi=1 \TIPI\else \TIPITOT\fi\fi


\let\0=\noindent\def\pagina{{\vfill\eject}}

\def\media#1{{\langle#1\rangle}}
\def\ie{\hbox{\it i.e.\ }}
\let\dpr=\partial\def\\{\hfill\break}

\def\*{\vglue0.3truecm}\let\0=\noindent
\let\==\equiv

\def\mbe{{\\*\hfill\hbox{\it
mbe\kern0.5truecm}}\vskip3.truept}
\def\1{{-1}}
\let\io=\infty \def\V#1{\,\vec#1}   
    \let\ig=\int

\def\tende#1{\,\vtop{\ialign{##\crcr\rightarrowfill\crcr
              \noalign{\kern-1pt\nointerlineskip}
              \hskip3.pt${\scriptstyle #1}$\hskip3.pt\crcr}}\,}
\def\otto{\,{\kern-1.truept\leftarrow\kern-5.truept\to\kern-1.truept}\,}
\def\fra#1#2{{#1\over#2}}

\global\newcount\numpunt
\def\XWPR{{\it a priori}}
\def\ap#1{\def\9{#1}{\if\9.\global\numpunt=1\else\if\9,\global\numpunt=2\else
\if\9;\global\numpunt=3\else\if\9:\global\numpunt=4\else
\if\9)\global\numpunt=5\else\if\9!\global\numpunt=6\else
\if\9?\global\numpunt=7\else\global\numpunt=8\fi\fi\fi\fi\fi\fi
\fi}\ifcase\numpunt\or{\XWPR.}\or{\XWPR,}\or
{\XWPR;}\or{\XWPR:}\or{\XWPR)}\or
{\XWPR!}\or{\XWPR?}\or{\XWPR\ \9}\else\fi}

\def\fiat{{}}

\def\V#1{{\underline#1}}
\def\2{{1\over2}}
\def\EE{{\cal E}}
\def\CC{{\cal C}}

\def\T#1{{#1_{\kern-3pt\lower7pt\hbox{$\widetilde{}$}}\kern3pt}}
\def\VV#1{{\underline #1}_{\kern-3pt
\lower7pt\hbox{$\widetilde{}$}}\kern3pt\,}
\def\W#1{#1_{\kern-3pt\lower7.5pt\hbox{$\widetilde{}$}}\kern2pt\,}

\def\NN{{\cal N}}

\def\lis{\overline}


\def\indica{\leaders \hbox to 0.5cm{\hss.\hss}\hfill}


\def\guida{\leaders\hbox to 1em{\hss.\hss}\hfill}

\def\qq{{\bf q}}
\let\h=\eta\let\x=\xi\def\ie{{\it i.e. }}   
\newtoks\footline \footline={\hss\tenrm\folio\hss}
\footline={\rlap{\hbox{\copy200}}\tenrm\hss \number\pageno\hss}


\def\annota#1#2{\footnote{${}^#1$}{\nota#2\vfill}}

\fiat
\def\qq{{\V q}}
\vglue0.5cm
\centerline{\dodici
Reversible Anosov diffeomorphisms and large deviations.}
\*
\centerline{Giovanni Gallavotti\annota{*}{\nota
Fisica, Universit\`a di Roma La Sapienza, P.le Moro 2, 00185, Roma
Italia.}}
\*\*
\0{\it Abstract: the volume contraction obeys a large deviation
rule.}\annota{1}{\nota Archived in $mp\_arc@math.utexas.edu$, \#
95-19}

\*\*
\0{\it\S1. -- Anosov maps and thermodynamic formalism.}
\numsec=1\numfor=1\*

This section reviews the basic results of Sinai's theory of transitive
Anosov diffeomorphisms. It is meant for the reader with some familiarity
with the subject and it should be used for notations only. Expert
readers will probably find the rest of the paper self contained, without
this section.

Let $\CC$ be a $d$--dimensional, $C^\io$, compact manifold and let $S$
be a $C^\io$, transitive Anosov diffeomorphism, [AA], [S]. If
$W^u_x,W^s_x$ denote the {\it unstable} or {\it stable} manifold at
$x\in\CC$, we call $W^{u,\d}_x,W^{s,\d}_x$ the connected parts of
$W^u_x,W^s_x$ containing $x$ and contained in the sphere with center $x$
and radius $\d$. Let $d_u,d_s$ be the {\it dimensions} of $W^u_x,W^s_x$:
$d=d_u+d_s$. It is convenient to take $\d$ always smaller than the
smallest curvature radius of $W^u_x,W^s_x$ for $x\in\CC$. Transitivity
means that $W^u_x,W^s_x$ are dense in $\CC$ for all $x\in \CC$.

The map $S$ can be regarded, locally near $x$, either as a map of $\CC$
to $\CC$ or of $W^u_x$ to $W^u_{Sx}$, or of $W^s_x$ to $W^s_{Sx}$.  The
{\it jacobian matrices} of the "three" maps will be $d\times d$,
$d_u\times d_u$ and $d_s\times d_s$ matrices denoted respectively $\dpr
S(x), \,\dpr S(x)_u,\,\dpr S(x)_s$.  The absolute
values of the respective determinants
will be denoted, respectively, $\L(x)$, $\L_u(x)$,
$\L_s(x)$
and are H\"older continuous functions, strictly positive (in fact
$\L(x)$ is $C^\io$), [S], [AA]. Likewise one can define the jacobians of
the $n$--th iterate of $S$ which are denoted by appending a label $n$
to $\L,\L_u,\L_s$ and are related to the latter by:

$$\eqalign{\L_n(x)=&\prod_{j=0}^{n-1}\L(S^jx),\quad\L_{u,n}(x)=
\prod_{j=0}^{n-1}\L_u(S^jx),\quad\L_{s,n}(S^jx)=
\prod_{j=0}^{n-1}\L_{s,n}(S^jx)\cr
\L_n(x)=&\L_{u,n}(x)\,\L_{s,n}(x)\chi_n(x)\cr}\Eq(1.1)$$

\0and $\chi_n(x)=\fra{\sin\a(S^nx)}{\sin\a(x)}$ is the ratio of the
sines of the {\it angles} $\a(S^nx)$ and $\a(x)$ between $W^u$ and $W^s$ at
the points $S^nx$ and $x$.  Hence $\chi_n(x)$ is bounded above and
below in terms of a constant $B>0$: $B^{-1}\le\chi_n(x)\le B$, for all
$x$ (by the transversality of $W^u$ and $W^s$).

A set $E$ is a {\it rectangle} with {\it center} $x$ and {\it axes}
$\D^u,\D^s$ if:\\
1) $\D^u,\D^s$ are two connected surface elements of $W^u_x,W^s_x$
containing $x$.\\
2) for any choice of $\x\in\D^u$ and $\h\in\D^s$ the local manfolds
$W^{s,\d}_\x$ and $W^{u,\d}_\h$ intersect in one and only one point
$x(\x,\h)=W^{s,\d}_\x\cap W^{u,\d}_\h$. The point $x(\x,\h)$ will also
be denoted $\x\times\h$.\\
3) the boundaries $\dpr\D^u$ and $\dpr\D^s$ (regarding the latter sets
as subsets of $W^u_x$ and $W^s_x$) have zero surface area on $W^u_x$ and
$W^s_x$.\\
4) $E$ is the set of points $\D^u\times\D^s$.

Note that {\it any} $x'\in E$ can be regarded as the center of $E$
because there are $\D^{\prime u},\D^{\prime s}$ both containing $x'$ and
such that $\D^u\times\D^s\= \D^{\prime u}\times\D^{\prime s}$.
Hence each $E$ can be regarded as a rectangle centered at any $x'\in E$
(with suitable axes). See figure.
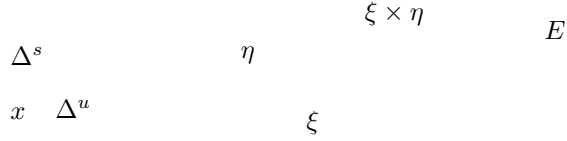
\figini{figurediffxxx}
\8</punto { gsave >
\8<3 0 360 newpath arc fill stroke grestore} def>
\8</puntino { gsave >
\8<2 0 360 newpath arc fill stroke grestore} def>
\8<>
\8</origine1assexper2pilacon|P_2-P_1| { >
\8<4 2 roll 2 copy translate exch 4 1 roll sub >
\8<3 1 roll exch sub 2 copy atan rotate 2 copy >
\8<exch 4 1 roll mul 3 1 roll mul add sqrt } def>
\8<>
\8</punta0{0 0 moveto dup dup 0 exch 2 div lineto 0 >
\8<lineto 0 exch 2 div neg lineto 0 0 lineto fill >
\8<stroke } def>
\8<>
\8</dirpunta{>
\8<gsave origine1assexper2pilacon|P_2-P_1| >
\8< 0 translate 7 punta0 grestore} def>
\8<>
\8</uno{
\8<dup mul div neg} def>
\8</due{
\8<sub div} def>
\8</p{
\8<dup 5 1 roll 3 -1 roll due 3 1 roll exch uno add} def>
\8<>
\8<gsave>
\8<40 40 40 0 360 arc>
\8<20 40 moveto 60 40 lineto>
\8<40 20 moveto 40 60 lineto>
\8<stroke>
\8<>
\8<140 40 40 0 360 arc >
\8<120 40 moveto 160 40 lineto>
\8<140 20 moveto 140 60 lineto>
\8<stroke>
\8<150 40 2 0 360 arc fill>
\8<140 50 2 0 360 arc fill>
\8<150 50 2 0 360 arc fill>
\8<stroke>
\8<175 75 moveto 160 60 lineto>
\8<175 75 160 60 dirpunta>
\8<150 10 moveto 150 70 lineto>
\8<110 50 moveto 165 50 lineto>
\8<stroke>
\8<>
\8<240 40 40 0 360 arc>
\8<220 20 moveto 220 60 lineto>
\8<260 60 lineto 260 20 lineto 260 20 220 20 lineto>
\8<220 40 moveto 260 40 lineto>
\8<240 20 moveto 240 60 lineto>
\8<stroke>
\8<220 40 puntino>
\8<260 40 puntino>
\8<240 60 puntino>
\8<240 20 puntino>

\8<grestore>
\figfin

\eqfig{260pt}{90pt}{
\ins{43pt}{37pt}{$x$}
\ins{43pt}{60pt}{$\D^s$}
\ins{60pt}{40pt}{$\D^u$}
\ins{155pt}{36pt}{$\x$}
\ins{130pt}{60pt}{$\h$}
\ins{177pt}{77pt}{$\x\times\h$}
\ins{245pt}{70pt}{$E$}
}{figurediffxxx}{}
\*
{\nota
\0The circle is a small neighborhood of $x$; the first picture shows the
axes; the intermediate picture shows the $\times$ operation and
$W^{u,\d}_\h, W^{s,\d}_\x$; the third picture shows the rectangle $E$
with the axes and the four marked points are the boundaries $\dpr\D^u$
and $\dpr\D^s$. The picture refers to a two dimensional case and the
stable and unstable manifolds are drawn as flat, \ie the $\D$'s are very
small compared to the curvature of the manifolds. The center $x$ is
drawn in a central position, but it can be {\it any} other point of $E$
provided $\D^u$ and $\D^s$ are correspondingly redefined. One should
meditate on the symbolic nature of the drawing in the cases of higher
dimension.\vfill}

The {\it unstable boundary} of a rectangle $E$ will be the set $\dpr_u
E=\D^u\times\dpr\D^s$; the {\it stable boundary} will be $\dpr_s
E=\dpr\D^u\times\D^s$.  The boundary $\dpr E$ is therefore $\dpr
E=\dpr_s E\cup\dpr_u E$.  The set of the {\it interior points}
of $E$ will be
denoted $E^0$.  A {\it pavement} of $\CC$ will be a covering
$\EE=(E_1,\ldots,E_\NN)$ of $\CC$ by $\NN$ rectangles with pairwise
disjoint interiors.  The {\it stable (or unstable) boundary} $\dpr_s\EE$
(or $\dpr_u \EE$) of $\EE$
is the union of the stable (or unstable) boundaries of the rectangles
$E_j$: $\dpr_u \EE=\cup_j\dpr_u E_j$ and
$\dpr_s \EE=\cup_j\dpr_s E_j$.

A pavement $\EE$ is called {\it markovian} if its stable boundary
$\dpr_s \EE$ retracts on itself under the action of $S$ and its unstable
boundary retracts on itself under the action of $S^{-1}$, [S], [Bo];
this means:

$$S\dpr_s\EE\subseteq \dpr_s\EE,\qquad S^{-1}\dpr_u\EE\subseteq \dpr_u\EE
\Eq(1.2)$$

\0Setting $M_{j,j'}=0$, $j,j'\in\{1,\ldots,\NN\}$, if $S E^0_j\cap
E^0_{j'}=\emptyset$ and $M_{j,j'}=1$ otherwise we call $C$ the set of
sequences $\V j=(j_k)_{k=-\io}^\io$, $j_k\in\{1,\ldots,\NN\}$ such that
$M_{j_k,j_{k+1}}\=1$.  The transitivity of the system $(\CC,S)$ implies
that $M$ is {\it transitive}: \ie there is a power of the matrix $M$
with all entries positive.  The space $C$ will be called the space of
the {\it compatible symbolic sequences}.  If $\EE$ is a markovian
pavement and $\d$ is small enough the map:

$$X: \V j\in C\,\to\,x=\bigcap_{k=-\io}^\io S^{-k} E_{j_k}\in\CC\Eq(1.3)$$

\0is continuous and $1-1$ between the complement $\CC_0\subset\CC$
of the set $N=
\cup_{k=-\io}^\io
S^k\dpr \EE $ and the complement $C_0\subset C$ of
$X^{-1}(N)$. This map is called the {\it
symbolic code} of the points of $\CC$: it is a code that associates with
each $x\not\in N$ a sequence of symbols $\V j$ which are the labels of
the rectangles of the pavement that are successively visited by the
motion $S^jx$.

The symbolic code $X$ transforms the action of $S$ into the {\it left shift}
$\th$ on $C$: $S X(\V j)= X(\th \V j)$. A key result, [S], is that it
transforms the {\it volume measure} $\m_0$ on $\CC$ into a {\it Gibbs
distribution}, [LR], [R], $\lis\m_0$ on $C$ with formal hamiltonian:

$$H(\V j)=\sum_{k=-\io}^{-1} h_-(\th^k\V j)+h_0(\V j)+\sum_{k=0}^\io
h_+(\th^k \V j)\Eq(1.4)$$

\0where, see \equ(1.1):

$$h_-(\V j)=-\log \L_s(X(\V j)),\quad h_+(\V j)=\log \L_u(X(\V j)),\quad
h_0(\V j)=\log\sin\a(X(\V j))\Eq(1.5)$$

If $F$ is smooth on $\CC$ the function $\lis F(\V j)=F(X(\V j))$ can
be represented in terms of suitable functions $\F_k(j_{-k},\ldots,j_k)$
as:

$$\lis F(\V j)=\sum_{k=1}^\io \F_k(j_{-k},\ldots,j_k),\qquad
|\F_k(j_{-k},\ldots,j_k)|\le \f e^{-\l k}\Eq(1.6)$$

\0where $\f>0,\l>0$. In particular $h_\pm$ (and $h_0$) enjoy the property
\equ(1.6) ({\it short range}).

If $\lis\m_+,\lis\m_-$ are the Gibbs states with formal hamiltonians:

$$\sum_{k=-\io}^\io h_+(\th^k\V j),\qquad
\sum_{k=-\io}^\io h_-(\th^k\V j)\Eq(1.7)$$

\0the distributions $\m_\pm$ on $\CC$, images of $\lis\m_\pm$ via the
code $X$ in \equ(1.3), will be the {\it forward} and {\it backward
statistics} of the volume distribution $\m_0$ (corresponding to
$\lis\m_0$ via the code $X$), [S]. This means that:

$$\lim_{T\to\io}\fra1T\sum_{k=0}^{T-1} F(S^{\pm k}x)=\ig_\CC \m_\pm(dy)
F(y)\=\m_\pm (F)\Eq(1.8)$$

\0for all
smooth $F$ and for $\m_0$--almost all $x\in\CC$. The distributions
$\m_\pm$ are often called the {\it SRB distributions}, [ER]; the above
statements and \equ(1.7),\equ(1.8) constitute the content of a well
known theorem by Sinai, [S].

An approximation theorem for $\m_+$ can be given in terms of the {\it
coarse graining} of $\CC$ generated by the markovian pavement
$\EE_T=\bigvee_{k=-T}^T S^{-k}\EE$.\annota{2}{\nota Where $\vee$
denotes the operation which, given two pavements $\EE,\EE'$ generates a
new pavement $\EE\vee\EE'$: the rectangles of $\EE\vee\EE'$ simply
consist of all the intersections $E\cap E'$ of pairs of rectangles
$E\in\EE$ and $E'\in\EE'$.} If
$E_{j_{-T},\ldots,j_T}\=\cap_{k=-T}^T S^{-k} E_{j_k}$ and if
$x_{j_{-T},\ldots,j_T}$ is a point, arbitrarily chosen, in the coarse
grain set $E_{j_{-T},\ldots,j_T}$ we define the distribution
$\m_{T,\t}$ by setting:

$$\eqalign{
\m_{T,\t}(F)\=&\ig_\CC \m_{T,\t}(dx) F(x)=
\fra{\sum_{j_{-T},\ldots,j_T}\lis\L_{u,\t}^{\,-1}
(x_{j_{-T},\ldots,j_T})
F(x_{j_{-T},\ldots,j_T})}{\sum_{j_{-T},
\ldots,j_T}\lis\L_{u,\t}^{\,-1}(x_{j_{-T},\ldots,j_T})}\cr
\lis\L_u(x){\buildrel def \over =}&
\prod_{k=-\t/2}^{\t/2-1}\L_u(S^kx)\cr}\Eq(1.9)$$

Then for all smooth $F$ it is: $\lim_{T\ge\t/2,\,\t\to\io}
\m_{T,\t}(F)=\m_+(F)$. Note that equation \equ(1.9) can also be
written:

$$\m_{T,\t}(F)=
\fra{\sum_{j_{-T},\ldots,j_T}e^{-\sum_{k=-\t/2}^{\t/2-1}
h_+(\th^k\V j^0)} F(X(\V j^0)}
{\sum_{j_{-T},
\ldots,j_T}
e^{-\sum_{k=-\t/2}^{\t/2-1}
h_+(\th^k\V j^0)}}\Eq(1.10)$$

\0where $\V j^0\in C$  is a compatible sequence, {\it arbitrarily
chosen}, which agrees with
$j_{-T},\ldots,j_T$ between $-T$ and $T$. (\ie $X(\V
j^0)=x_{j_{-T},\ldots,j_T}\in E_{j_{-T},\ldots,j_T}$).
\*
\0{\it Notation:} to simplify the notations we shall write $\qq$ for
the elements $\qq=(j_{-T},\ldots,j_T)$ of $\{1,\ldots,N\}^{2T+1}$;
and $E_\qq$ will denote $E_{j_{-T},\ldots,j_T}$ and $x_\qq$ a point of
$E_\qq$.
\*
\0{\it Remark:} Note that the weights in \equ(1.10) depend on the
special choices of the centers $x_\qq$ (\ie of $\V j^0$); but if $x_\qq$
varies in $E_\qq$ the weight of $x_\qq$ changes by at most a factor,
bounded above by some $B<\io$ and below by $B^{-1}$, for all $T\ge0$:
this is a consequence of the short range properties of $h_+$ expressed
by \equ(1.6)).
\*
The last formula shows that the forward statistics of $\m_0$ can be
regarded as a Gibbs state for a {\it short range one dimensional spin
chain with a hard core interaction}. The spin at $x$ is the value of
$j_x\in\{1,\ldots,\NN\}$; the short range refers to the fact
that, $\L_u(x)$ being smooth, the function $h_+(\V j)\=\log \L_u(X(\V
j))$ can be represented as in \equ(1.6) where the $\F_k$ play he role of
"many spins" interaction potentials and the hard core refers to the fact
that the only spin configurations $\V j$ allowed are those with
$M_{j_k,j_{k+1}}\=1$ for all integers $k$.

\vskip0.5cm
\0{\it\S2. -- Reversible dissipative systems and results.}
\numsec=2\numfor=1\*

Let $(\CC,S)$ be a transitive, smooth Anosov system, see \S1, and let
$\L(x)=|\det\dpr S(x)|$; let $\m_\pm$ be the forward and backward
statistics of the volume measure $\m_0$ (\ie the SRB distributions for
$S$ and $S^{\,-1}$).\*

\0{\it Definition (A): The system $(\CC,S)$ is {\sl dissipative} if:

$$-\ig_{\CC}\m_\pm(dx)\log\L(x)^{\pm1}=\s_\pm>0\Eq(2.1)$$}
\*
\0{\it Definition (B): The system $(\CC,S)$ is {\sl reversible} if
there is an isometric involution $i:\CC\otto\CC$, ($i^2=1$), such that:
$i S= S^{\,-1} i$.
\*\penalty10000
We consider from now on only dynamical systems $(\CC,S)$ verifying
(A),(B).}
\*\penalty-200
\0{\it Remarks:} \\
1) (A),(B) imply $\s_+=\s_-$ and $\L(x)=\L(ix)^{\,-1}$;
furthermore $iW^u_x=W^s_{ix}$ and the dimensions of the stable and
unstable manifolds $d_s,d_u$ are equal: $d_u=d_s$ and $d=d_u+d_s$ is
even.\\
2) if $\L_u(x),\L_s(x)$ denote the absolute values of the jacobian
determinants of $S$ as a map of $W^u_x$ to $W^u_{Sx}$ and of
$W^s_x$ to $W^s_{Sx}$, then $\L_u(x)=\L_s(ix)^{\,-1}$.
\*
\penalty-1000
\0{\it Definition: The {\sl dimensionless entropy production rate} or
the {\sl phase space contraction rate} at $x\in\CC$ and
over a time $\t$ is the function $\e_\t(x)$:

$$x\to\e_\t(x)=\fra1{\s_+\t}\sum_{j=-\t/2}^{\t/2-1}\log
\L^{\,-1}(S^jx)=\fra1{\s_+\t}\log \lis\L^{\,-1}(x)\Eq(2.2)$$

\0with $\lis\L(x){\buildrel def \over =}\prod_{-\t/2}^{\t/2-1}
\L(S^jx)$ (see \equ(1.9)).}
\*

\0{\it Remarks:}

i) by definition (see (2.1)):

$$\media{\e_\t}_+=\lim_{T\to+\io}
\fra1T\sum_{j=0}^{T-1}\e_\t(S^jx) \=\ig\m_+(dy)\e_\t(y)=1\Eq(2.3)$$

\0with $\m_0$--probability $1$.

ii) Note that, on the other hand, $\lim_{\t\to\io}\e_\t(x)=0$ with
$\m_0$--probability $1$, by the reversibility (B).
\*
Here we prove the following {\it fluctuation theorem}:
\*
\penalty10000
\0{\it Fluctuation theorem: There exists $p^*>0$ such that
the SRB distribution $\m_+$ verifies:

$$p-\d\le
\lim_{\t\to\io}\fra1{\s_+\t}\log\fra{\m_+(\{\e_\t(x)\in[p-\d,p
+\d]\})}{\m_+(\{\e_\t(x)\in-[p-\d,p+\d]\})}\le p+\d\Eq(2.4)$$

\0for all $p, \,|p|<p^*$.}
\*
The above theorem was first informally proved in [GC1],[GC2] where its
interest for nonequilibrium statistical mechanics was pointed out.
Although I think that the physical interest of the theorem far
outweighs its mathematical aspects, see also [G1], it appears that it
might be useful to write the explicit and formal proof (more detailed
than [G2]) described below.  The theorem can be regarded as a large
deviation result for the probability distribution $\m_+$.

The strategy that I shall follow to take advantage of the existing
literature is the following. First the function \equ(2.2) is converted
to a function on the spin configurations $\V j\in C$ (see \S1):

$$\tilde\e_\t(\V j)=\e_\t(X(\V
j))=\fra1{\t}\sum_{k=-\t/2}^{\t/2-1} L(\th^k\V j)\Eq(2.5)$$

\0where $L(\V j)\=\fra1{\s_+}\log \L(X(\V j))^{\,-1}$ has a {\it short range}
representation of the type \equ(1.6).

The SRB distribution $\m_+$ can be regarded as a Gibbs state $\lis\m_+$
with short range potential on the space $C$ of the compatible
symbolic sequences, associated with a Markov partition $\EE$. Therefore
there is a function $\z(s)$ real analytic in $s$ for $s\in(-p^*,p^*)$
for a suitable $p^*>0$, strictly convex and such that if
$p<p^*$ and $[p-\d,p+\d]\subset(-p^*,p^*)$ it is:

$$\fra1\t \log\lis\m_+(\{\tilde\e_\t(\V j)\in[p-\d,p+\d]\})\tende{\t\to\io}
\max_{s\in[p-\d,p+\d]}-\z(s)\Eq(2.6)$$

\0and the difference between the r.h.s.  and the l.h.s.  tends to $0$
bounded by $D\t^{\,-1}$ for a suitable constant $D$.  The function
$\z(s)$ is the Legendre transform of the function $\l(\b)$ defined as:

$$\l(\b)=\lim_{\t\to\io}\fra1\t\log \ig e^{\t\b \tilde\e_\t(\V j)}\,
\lis\m_+(d\V j)\Eq(2.7)$$

\0\ie $\l(\b)=\max_{s\in(-p^*,p^*)}(\b s-\z(s))$, where $p^*$
can be taken $p^*=\lim_{\b\to+\io}\b^{-1}\l(\b)$ and the function
$\l(\b)$ is a real analytic, [CO], strictly convex function of
$\b\in(-\io,\io)$ and asymptotically linear:
$\b^{-1}\l(\b)\tende{\b\to\pm\io}\pm p^*$.

The above \equ(2.6) is a "large deviations theorem" for one dimensional
spin chains with short range interactions, [L].

Hence it will be sufficient to prove the following; if
$I_{p,\d}=[p-\d,p+\d]$:

$$\fra1{\s_+\t}
\log\fra{\lis\m_+(\{\tilde\e_\t(x)\in I_{p,\d\pm\h(\t)}\})}{\lis\m_+
(\{\tilde\e_\t(x)\in I_{-p,\d\mp\h(\t)})\}}\cases{<  p+\d+\h'(\t)\cr
>p-\d-\h'(\t)\cr}\Eq(2.8)$$

\0with $\h(\t),\h'(\t)\tende{\t\to\io}0$.
\*
\0{\it\S3. -- Thermodynamic formalism (proof the fluctuation theorem).}
\numsec=3\numfor=1
\*

Let, for $n$ odd, $\V j_X=(j_x,j_{x+1},\ldots,j_{x+n-1})$ if
$X=(x,{x+1},\ldots,{x+n-1})$, and call $\lis X=x+(n-1)/2$ the
{\it center} of $X$. If $\V j\in C$ is an infinite spin configuration we
also denote $\V j_X$ the set of the spins with labels $x\in X$.
The left shift of the interval $X$ will be denoted by $\th$; \ie
by the same symbol of the left shift of a (infinite) spin configuration
$\V j$.

Let $l_X(\V j_X)=l^{(n)} (j_x,j_{x+1},\ldots,j_{x+n-1})$, $h^+_X(\V
j_X)=h^{(n)}_+(j_x,j_{x+1}, \ldots,j_{x+n-1})$ be translation invariant
functions, \ie functions such that $l_{\th X}(\V j)\= l_X(\V j)$ and
$h^+_{\th X(\V j) }=h^+_X(\V j)$, and such that the functions $h_+(\V
j)$, see \equ(1.6), and $L(\V j)$, see \equ(2.5), can be written:

$$\eqalign{
L(\V j)=&\sum_{\lis X= 0}l_X(\V j_X), \qquad h_+(\V j)=\sum_{\lis X=
0}h^+_X(\V
j_X)\cr
|l_X(\V j_X)|\le& b_1 e^{-b_2 n},\qquad\kern1cm |h^+_X(\V j_X)|\le b e^{-b'
n}\cr}\Eq(3.1)$$

\0for suitable constants $b_1,b_2,b,b'$. Then $\t\tilde\e_\t (\V
j)$ can be written as $\sum_{\lis X\in[-\t/2,\t/2-1]} l_X(\V j_X)$.

Hence $\t\tilde\e_\t(\V j)$ can be approximated by $\t\tilde \e_\t^M(\V j)
={\sum}^{(M)} l_X(\V j_X)$ where $\sum^{(M)}$ means summation over the sets
$X\subseteq[-\fra12\t-M,\fra12\t+M]$, while
$\lis X$ is in $[-\fra12\t,\fra12\t-1]$. The approximation is described
by:

$$|\t\tilde\e^M_\t(\V j)-\t\tilde\e_\t(\V j)|\le b_3 e^{-b_4
M}\=\h_0(M)\Eq(3.2)$$

\0for suitable\annota{3}{\nota One can check from \equ(3.1),
that the constants $b_3,b_4$ can be expressed as simple functions of
$b_1,b_2$.} $b_3,b_4$ and for all $M\ge0$. Therefore if
$I_{p,\d}=[p-\d,p+\d]$ and $M=0$ we have:

$$\m_+(\{\e_\t(x)\in I_{p,\d}\})\cases{\le \lis\m_+(\{\tilde\e^0_\t\in
I_{p,\d+b_3/\t}\})\cr \ge \lis\m_+(\{\tilde\e^0_\t\in
I_{p,\d-b_3/\t}\})\cr}\Eq(3.3)$$

It follows from the general theory of $1$--dimensional Gibbs
distributions, [R2], that the $\lis\m_+$--pro\-ba\-bi\-li\-ty of a spin
configuration coinciding with $\V j_{[-\t/2,\t/2]}$ in the interval
$[-\fra12 \t,\fra12\t]$,\annota{4}{\nota \ie the spin
configurations $\V j'$ such that $j'_x=j_x$,
$x\in[-\fra12\t,\fra12\t]$.} is:

$$\fra{\Big[e^{-{\sum}^*h^+_X(\V j_X)}\Big]}
{\sum_{\V j'_{[-\t/2,\t/2]}}\Big[\cdot\Big]}\,
P(\V j_{[-\t/2,\t/2]})\Eq(3.4)$$

\0where $\sum^*$ denotes summation over all the $X\subset
[-\t/2,\t/2-1]$; the denominator is just the sum of terms like the
numerator, evaluated at a generic (compatible) spin configuration $\V
j'_{[-\t/2,\t/2]}$; finally $P$ verifies the bound, [R2]:

$$B_1^{-1}< P(\V j_{\,[-\t/2,\t/2]})< B_1\Eq(3.5)$$

\0with $B_1$ a suitable constant independent of $\V j_{\,[-\t/2,\t/2]}$
and of $\t$ ($B_1$ can be explicitly estimated in terms of $b,b'$).
Therefore from \equ(3.3) and \equ(3.4) we deduce for any $T\ge\t/2$:

$$\eqalign{
&\m_+(\e_\t(x)\in I_{p,\d})\le \lis\m_+(\tilde
\e^0_\t\in I_{p,\d+b_3/\t})\le\cr
&\le B_2\,\m_{T,\t}(\tilde\e^0_\t\in I_{p,\d+b_3/\t})\le
B_2\,\m_{T,\t}(\tilde\e_\t\in I_{p,\d+2b_3/\t})\cr}\Eq(3.6)$$

\0for some constant $B_2>0$; and likewise a lower bound is obtained by
replacing $B_2$ by $B_2^{-1}$ and $b_3$ by $-b_3$.

Then if $p<p^*$ and $I_{p,\d}\subset (-p^*,p^*)$ the set of the
rectangles $E\in\bigvee_{-T}^T S^{-k}\EE$ with center $x$
such that $\e_\t(x)\in I_{p,\d}$
is {\it not empty}.

We immediately deduce the lemma:
\*
\0{\it Lemma 1: the distributions $\m_+$ and $\m_{T,\t}$, $T\ge\fra12\t$,
verify:

$$\fra1{\t\s_+}\log \fra{\m_+(\e_\t(x)\in I_{p,\d\mp 2b_3/\t})}
{\m_+(\e_\t(x)\in- I_{p,\d\pm 2b_3/\t})}\  \cases{
<\fra{\log B_2^2}{\t\s_+}+\fra1{\t\s_+}\log
\fra{\m_{T,\t}(\tilde\e_\t\in I_{p,\d})}{\m_{T,\t}(\tilde\e_\t\in- I_{p,\d})}
\cr
> -\fra{\log B_2^2}{\t\s_+}+\fra1{\t\s_+}\log
\fra{\m_{T,\t}(\tilde\e_\t\in I_{p,\d})}{\m_{T,\t}(\tilde\e_\t\in-
I_{p,\d})}\cr}\Eq(3.7)$$
\penalty10000
\0for $I_{p,\d}\subset [-p^*,p^*]$ and for $\t$ so large that
$p+\d+2b_3/\t< p^*$.}
\penalty10000
\*
Hence \equ(2.8) will follow if we can prove:
\vfill\pagina
\*
\0{\it Lemma 2: there is a constant $\lis b$
such that the approximate SRB distribution $\m_{T,\t}$ verifies:

$$\fra1{\s_+\t}\log
\fra{\m_{T,\t}(\tilde\e\in I_{p,\d})}{\m_{T,\t}(\tilde\e\in-
I_{p,\d})}\ \cases{\le p+\d+ \lis b/\t\cr
\ge p-\d -\lis b/\t\cr}\Eq(3.8)$$

\0for $\t$ large enough (so that $\d+\lis b/\t<p^*-p$) and for all
$T\ge\t/2$.}

\*
The latter lemma will be proved in \S4 and it is the only statement that
does not follow from the already existing literature.

\vskip0.5cm
\0{\it\S4. -- Time reversal symmetry and large deviations.}
\numsec=4\numfor=1\*

The relation \equ(3.7) holds for any choice of the Markov partition
$\EE$. Note that if $\EE$ is a Markov pavement also $i\EE$ is such
(because $iS=S^{-1}i$ and $i W^u_x=W^s_{ix}$); furthermore if $\EE_1$
and $\EE_2$ are Markov pavements also $\EE=\EE_1\vee\EE_2$ are markovian.
Therefore:

\*\0{\it Lemma 3: there exists a time reversal Markov pavement $\EE$,
\ie a Markov pavement such that $\EE=i\EE$.}

\*
This can be seen by taking any Markov pavement $\EE_0$ and setting
$\EE=\EE_0\vee i\EE_0$. Alternatively one could construct the Markov
pavement in such a way that it verifies automatically the symmetry [G2].
Since the center of a rectangle $E_\qq\in\EE_T$ can be taken to be any
point $x_\qq$ in the rectangle $E_\qq$ we can and shall suppose that the
centers of the rectangles in $\EE_T$ have been so chosen that the center
of $i E_\qq$ is $i x_\qq$, \ie the time reversal of the center $x_\qq$
of $E_\qq$.

For $\t$ large enough the set of configurations $\qq=\V j_{\,[-T,T]}$
such that $\e_\t(x)\in I_{p,\d}$ for all $x\in E_\qq$ is not
empty\annota{5}{\nota Note that $p^*=\sup_x \limsup_{\t\to+\io}
\e_\t(S^{\t/2}x)$ and let $p\in(-p^* +\d,p^*-\d)$; furthermore $\z(s)$
is smooth, hence $>-\io$, for all $|s|<p^*$.} and the ratio in
\equ(3.8) can be written, if $x_\qq$ is the center of $E_\qq
\in\bigvee_{-T}^T\EE$, as:

$$\fra{\sum_{\e_\t(x_\qq)\in I_{p,\d}}
\lis\L^{\,-1}_{u,\t}(x_\qq)}{\sum_{\e_\t(x_\qq)
\in- I_{p,\d}}\lis\L^{\,-1}_{u,\t} (x_\qq)}=
\fra{\sum_{\e_\t(x_\qq)\in I_{p,\d}}
\lis\L^{\,-1}_{u,\t}(x_\qq)}{\sum_{\e_\t(x_\qq)\in
I_{p,\d}}\lis\L^{\,-1}_{u,\t}(i\,x_\qq)}\Eq(4.1)$$

But the time reversal symmetry implies that
$\lis\L_{u,\t}(x)=\lis\L_{s,\t}^{\,-1}(i x)$, see remark 2) following
definition (B), \S2.\annota{6}{\nota here it is essential that
$\lis\L_{u,\t}(x)$ is the expansion of the unstable manifold between the
initial point $S^{-\t/2}x$ and the final point $S^{\t/2}x$ which is a
trajectory of time length $\t$, which at its central time is in $x$.}
This permits us to change \equ(4.1) into:

$$\fra{\sum_{\e_\t(x_\qq)\in I_{p,\d}}
\lis\L_{u,\t}^{\,-1}(x_\qq)}{\sum_{\e_\t(x_\qq)\in
I_{p,\d}}\lis\L_{s,\t}(i\,x_\qq)}\ \cases{< \max_\qq
\lis\L^{\,-1}_{u,\t}(x_\qq)
\lis\L^{\,-1}_{s,\t}(x_\qq)
\cr>\min_\qq
\lis\L^{\,-1}_{u,\t}(x_\qq)
\lis\L^{\,-1}_{s,\t}(x_\qq)\cr}
\Eq(4.2)$$

\0where the maxima are evaluated as $\qq$ varies with $\e_\t(x_\qq)\in
I_{p,\d}$.
\\\hbox{}\kern0.3cm
By \equ(1.1) we can replace
$\lis\L^{\,-1}_{u,\t}(x)\lis\L^{\,-1}_{s,\t}(x)$ with
$\lis\L_\t^{-1}(x)B^{\pm1}$, see \equ(1.9), \equ(2.2); thus
noting that by definition of the set of $\qq$'s in the maximum operation
in \equ(4.2) it is $\fra1{\s_+\t}\log \lis\L^{\,-1}_\t(x_\qq)
\in I_{p,\d}$, we see that \equ(3.8) follows with $\lis b
=\fra1{\s_+\t}\log B$.

\*
\0{\it Corollary: the above analysis gives us a concrete bound on the
speed at which the limits in \equ(2.4) are approached. Namely the error
has order $O(\t^{-1})$.}
\*
\0This is so because the limit \equ(2.6) is reached at speed
$O(\t^{-1})$; furthermore the regularity of $\l(s)$ in \equ(2.6) and the
size of $\h(\t),\h'(\t)$ and the error term in
\equ(3.8) have all order $O(\t^{-1})$.

The above analysis proves a large deviation result for the probability
distribution $\m_+$: since $\ m_+$ is a Gibbs distribution, see
\equ(1.7), various other large deviations theorems hold fo them, [DV],
[E], [O], but unlike the above they are not related to the time reversal
symmetry.
\*
{\it Acknowledgements}: I am indebted to E.G.D. Cohen and J.L. Lebowitz
for many discussions and for their encouragement. This work is part of
the research program of the European Network on: "Stability and
Universality in Classical Mechanics", \# ERBCHRXCT940460.
\*
\0{\it References.}
\*
[AA] Arnold, V., Avez, A.: {\it Ergodic problems of classical
mechanics}, Benjamin, 1966.

[CO] Cassandro, M., Olivieri, E.: {\it Renormalization group and
analyticity in one dimension. A proof of Dobrushin's theorem},
Communications in mathematical physics, {\bf 80}, 255--269, 1981.

[E] Ellis, R.S.: {\it Entropy, large deviations and statistical mechanics},
New York, Springer Verlag, 1985.

[DV] Donsker, M.D., Varadhan, S.R.S.: {\it Asymptotic evolution of certain
Markov processes expectations for large time}, Communications in Pure
and Applied Mathematics, {\bf 28}, 279--301, 1975; {\bf 29}, 389--461,
1976; {\bf 36}, 182--212, 1983.

[G1] Gallavotti, G.  {\it Ergodicity, ensembles and irreversibility in
Boltzmann and beyond}: {\it mp$\_$arc @ math.  utexas.  edu}, \# 94-66;
to appear in J.  Statistical Physics,

[G2] Gallavotti, G. {\it Topics on chaotic dynamics},
{\it mp$\_$arc @ math. utexas. edu}, \# 94-333.

[G3] Gallavotti, G.: {\it Coarse graining in chaotic systems},
in preparation.

[GC1] Gallavotti, G., Cohen, E.G.D: {\it Nonequilibrium stationary
states}, {\it mp$\_$arc @ math. utexas. edu}, \# 94-340.

[GC2] Gallavotti, G., Cohen, E.G.D: {\it Dynamical nonequilibrium
ensembles}, in preparation.

[L] Lanford, O.: {\it Entropy and equilibrium states in classical
statistical mechanics}, ed. A. Lenard, Lecture notes in Physics,
Springer Verlag, vol. {\bf 20}, p. 1--113, 1973.

[O] Olla, S.: {\it Large deviations for Gibbs random fields},
Probability Theory and related fields, {\bf 77}, 343--357, 1988.

[R1] Ruelle, D.: {\it A measure associated with axiom A
attractors}, American Journal of Mathematics, {\bf98}, 619--654, 1976.

[R2] Ruelle, D.: {\it Statistical mechanics of a one dimensional
lattice gas}, Communications in Mathematical Physics, {\bf 9}, 267--278,
1968.

[S] Sinai, Y.: {\it Markov partitions and $C$-diffeomorphisms},
Functional Analysis and Applications, {\bf2}, 64--89, n.1 (engl. p. 61),
1968.  And: {\it Construction of Markov partitions}, Functional analysis
and Applications, {\bf2}, 70--80 n.2 (engl. p. 245), 1968.

\bye